# QUANTIFYING DISTRIBUTION SYSTEM RESILIENCE FROM UTILITY DATA: LARGE EVENT RISK AND BENEFITS OF INVESTMENTS
## AUTHOR PREPRINT JULY 20 2024

Arslan AHMAD
Iowa State University – USA
arslan@iastate.edu

Ian DOBSON
Iowa State University – USA
dobson@iastate.edu

## ABSTRACT

*We focus on large blackouts in electric distribution systems caused by extreme winds. Such events have a large cost and impact on customers. To quantify resilience to these events, we formulate large event risk and show how to calculate it from the historical outage data routinely collected by utilities' outage management systems. Risk is defined using an event cost exceedance curve. The tail of this curve and the large event risk is described by the probability of a large cost event and the slope magnitude of the tail on a log-log plot.*

*Resilience can be improved by planned investments to upgrade system components or speed up restoration. The benefits that these investments would have had if they had been made in the past can be quantified by "rerunning history" with the effects of the investment included, and then recalculating the large event risk to find the improvement in resilience. An example using utility data shows a 12% and 22% reduction in the probability of a large cost event due to 10% wind hardening and 10% faster restoration respectively.*

*This new data-driven approach to quantify resilience and resilience investments is realistic and much easier to apply than complicated approaches based on modeling all the phases of resilience. Moreover, an appeal to improvements to past lived experience may well be persuasive to customers and regulators in making the case for resilience investments.*

## INTRODUCTION

Overhead distribution systems are vulnerable to extreme wind. For example, the August 2020 upper midwest USA derecho caused ~11 billion dollars of damage and left more than one million customers without power. Moreover, extremes of weather are gradually increasing [1]. This motivates quantifying the resilience risk to distribution systems of extreme winds, as well as quantifying the benefits of planned investments to reduce these risks, and finding ways to help justify these investments to customers and regulators.

Almost all of the literature quantifying distribution system resilience either optimizes expected (mean) losses or addresses the resilience performance curves of specific extreme events, or uses reliability indices such as SAIDI that address system reliability averaged over the year [1]. Expected or average losses are dominated by more routine outages and do not directly measure extreme event risk. The field is starting to move beyond specific resilience events and average metrics. For example, Carrington [2] extracts resilience events of all sizes from observed data and obtains the overall statistics of resilience metrics from the outage and restore processes. Moreover, papers led by Dubey [3, 4] have pioneered simulation models that assess Value at Risk (VaR) and Conditional Value at Risk (CVaR) resilience metrics that directly quantify the risk of large events. While almost all resilience quantification in distribution systems uses detailed models of a subset of resilience processes to simulate and assess resilience [1], excellent opportunities are opening up to assess distribution system resilience directly from observed utility data. Ahmad [5] uses utility data not only to quantify resilience with metrics but to "rerun history" with the effects of investments in resilience included to quantify the benefits of those investments. However Ahmad [5] uses metrics for resilience events such as number of outages, duration, and customer hours not served, and does not use a metric directly describing risk.

In this workshop paper, we aim to:
1. Formulate new metrics that use utility data to quantify distribution system resilience in terms of the risk of large events, and
2. Extend the historical rerun method to quantify the effects of resilience investments on the large event risk.

The resilience investments that we consider are hardening poles by increasing their wind rating and faster restoration.

The historical rerun method quantifies the resilience improvement that a proposed resilience investment would have had if the investment had been made in the past [5]. Since it is driven by real data, this has the advantage of incorporating all the factors affecting resilience over the past period such as weather, trees, human factors, operating procedures, equipment aging, system reconfigurations, and restoration practices. Thus the historical rerun method has no modeling error from these factors. The historical rerun method does not predict the future, but the model-based methods of predicting the future with simulation must represent considerable complexities of all the phases of resilience and are very complicated, whereas the historical rerun method is driven by data and is much more simple and straightforward. Moreover, in communicating the benefits of a proposed resilience investment to stakeholders, the historical rerun method has some advantages: The benefits that would have applied to the lived experience of stakeholders in the past, both for particular large events and in general, may well be more persuasive than the benefits that are modeled and simulated for predicted events at some indeterminate time in the future.

## OUTAGE DATA AND EXTRACTING EVENTS

We use six years of detailed outage data recorded by a US distribution utility in this work. The dataset contains records of 32278 individual power outages that occurred in the utility's network. Each outage entry corresponds to an outage of a component in the distribution system and includes the number of customers affected during the outage, the outage's start and end time, and its cause codes. We exclude the scheduled and planned outages and only consider the unscheduled outages in this analysis.

To conduct the wind resilience investments analysis, we use NOAA weather data from weather stations available within the distribution network's geographic area. For each outage, we use the weather data from the closest available weather station. The overall distribution network is thus divided into multiple small areas based on the number of weather stations. More details on this are available in [5]. For this paper, we use one of the areas with 12715 unscheduled distribution system outages of at least 1-minute duration.

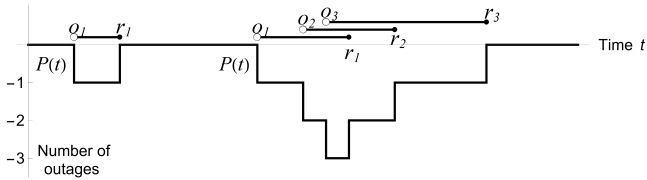

*Fig. 1: Event with one outage and an event with 3 outages [5, licensed under CC by 4.0]. Each outage's start time (open circle) and restore time (dot) are shown above the time axis. Below the time axis is the performance curve P(t) for each event. The event ends when P(t) returns to zero.*

We group the outages into 3706 resilience events during data pre-processing. Resilience events are formed by overlapping outages. The start of an event is defined by an initial outage that occurs when all components in the distribution system are operational, and the end of the same event is defined by the first subsequent time when all the components are restored. Two example events are shown in Fig. 1. More details about resilience events and their automatic extraction from the outage data are available in [5].

## ESTIMATING CUSTOMER COST

The customer cost of a power outage event can be described in terms of the total customer hours lost in that event as:

$$C = \beta A_{event}$$

where $A_{event}$ is the area under the customer performance curve of an outage event, which is equal to the total customer hours lost in that event [6], and $\beta$ is the average cost per customer per hour of an outage. The value of $\beta$ can be estimated in various ways such as customer surveys or online tools like DOE's ICE calculator and NREL's CDF calculator. We use $\beta$ = \$370.2 (2022 USD), based on the average proportions of customer classes (residential, commercial, industrial) in the utility and expert feedback from another utility.

The cost of a power outage to customers depends on different factors. These include the number of customers affected by the outage, outage duration, customer class, the affected customer's power outage risk level (houses with patients are at elevated risk), the criticality of services offered by the affected customer (hospitals, old homes, police, etc.), along with various other direct and indirect socio-economic factors. Incorporating all of these factors would give a more accurate yet complex model for the cost to customers. Different values of $\beta$ can be used for different customer classes and multiplied with each outage individually as per its affected customer class to get a more accurate estimate of the customer cost.

While we address here a resilience event's cost to customers, there are costs to the utility as well, which can be similarly modeled and analyzed to gain different insights.

## ESTIMATING LARGE EVENT RISK

One basic definition of risk associates probabilities with costs of events or groups of events [7], and can be described by the probability distribution of the cost. One useful way to present the probability distribution of cost is the cost exceedance function $\bar{F}_C(c) = P[C > c]$, which is the probability of the event cost $C$ exceeding the value $c$ as $c$ varies. The cost exceedance function is also known as the survival function or complementary cumulative distribution function (CCDF) or risk curve of $C$. Fig. 2 shows the customer cost exceedance function per event, obtained from the utility data.

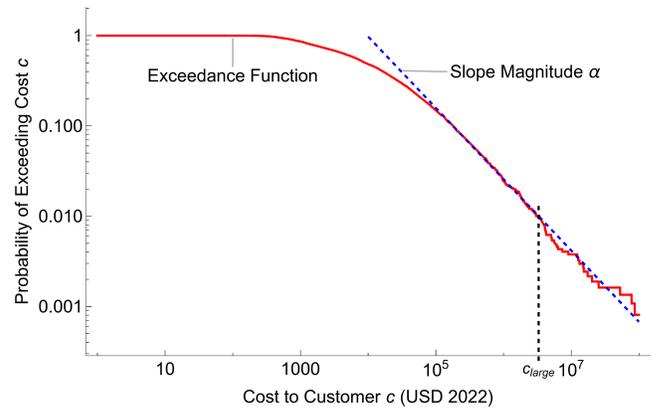

*Fig. 2: Customer event cost exceedance function, fitted tail distribution, and large-cost threshold on a log-log scale.*

**Probability of a large cost event**

To help describe and communicate the large event risk in Fig. 2, we define large cost events as those events with cost $C > c_{large}$, where $c_{large}$ is the threshold for the minimum large cost. Then the probability of a large cost event is the probability of an event cost exceeding $c_{large}$:

$$p_{large} = P[C > c_{large}]$$

The large cost threshold $c_{large}$ can be chosen by the utility, as long as there are sufficient large events to get a reasonable estimate of $p_{large}$. For our data we choose $c_{large} = 3.3 \times 10^6$ USD, which corresponds to the 99th percentile of the observed costs. It is shown as the vertical dotted black line on Fig. 2. It gives $p_{large} = 0.010$.

The large event cost probability can also be expressed in the form of an annual frequency of large-cost events $f_{large}$ by multiplying it by the average annual event rate $\bar{r}_{event}$:

$$f_{large} = p_{large}\, \bar{r}_{event}$$

For our utility data, $\bar{r}_{event} = 618$ events per year so that $f_{large} = 6.18$ events per year.

**Slope of the event cost exceedance curve**

The plot of the event cost exceedance curve shown in Fig. 2 exhibits a straight-line behavior after approximately $x_{min} = 10^5$. Since Fig. 2 has a log-log scale, this tail has an approximate power-law behavior given by:

$$\bar{F}(x) = \left(\frac{x}{x_{min}}\right)^{-\alpha}, \quad x > x_{min} \quad (1)$$

(To verify the straight-line behavior on the log-log plot, take the log of (1) to obtain $\log \bar{F}(x) = -\alpha \log x + \alpha \log x_{min}$.) The two parameters of such a power law distribution are the slope magnitude $\alpha$ of the power law fit and the cutoff $x_{min}$ after which the power law behavior starts. We estimate $\alpha = 0.789$ and $x_{min} = 130251$ USD for the utility cost data using the method of [8]. The portion of the customer event exceedance function to the right of $c_{large}$ defines the large event risk and its form can be reproduced just by using $\alpha$ and $p_{large}$.

Note that a *larger* value of slope magnitude $\alpha$ gives a steeper tail and *improved* resilience. $\alpha$ is the magnitude of the slope of the event cost exceedance function or CCDF of $C$ for large costs. It follows that the corresponding large cost slope magnitude of the probability density function of $C$ on a log-log plot is $\alpha+1$.

The probability $p_{large}$ that an event has a large cost can be compared to Value at Risk VaR [3]: For $p_{large}$ one fixes the threshold cost $c_{large}$, and then $p_{large}$ is the probability that the cost of an event exceeds $c_{large}$. For VaR one fixes a probability $p$ and then VaR is the minimum cost $c$ such that the event cost over a fixed period of time, such as one year, exceeds $c$ is $p$. So $p_{large}$ differs from VaR, but they both encode information about an event cost exceedance curve or an annual cost exceedance curve.

There are uncertainties in fitting the tails of heavy-tailed distributions, such as the cost exceedance curve [8, 9], including whether it is decisively fit by the power law (1) and the extent to which it can be reasonably extrapolated beyond the maximum cost event observed. This should be pursued in future work, and here we only indicate why the nature of the tail is so important by discussing the consequences if the tail is indeed best represented by the power law (1). Since for our utility data the estimated slope magnitude $\alpha$ is less than one, the tail is sufficiently heavy that taking the mean of costs involving the tail does not work because the mean of (1) is infinite. The consequent expected occurrence of events with extremely large costs suggests that there is no usable notion of an average or mean large event, such as used in Conditional Value at Risk CVaR. That is, some ways of quantifying the large event risk involving the mean large event cost may not be viable. Our two new metrics of large event risk are chosen to avoid this problem. Some of the issues posed by heavy-tailed probability distributions are well recognized in distribution systems (for example, the discussion of catastrophic events in [10, section 6.3]. Further careful analysis of catastrophic events and heavy tails is indicated.

Some of the previous work on distribution system resilience focuses on specific extreme events such as particular hurricanes. This work is valuable, but doesn't allow for a risk analysis since one cannot compute probabilities of large events without examining all events. Our approach uses observed events of all sizes to enable a risk analysis. Similarly, but working with simulations of a resilience model, Poudel [3] samples from all wind speeds to simulate the distribution system resilience and calculate risk metrics.

Much of the previous work considers the probability of component failure without considering costs, or uses reliability metrics such as SAIDI that measure average reliability over a year that, even if extreme events are included in the calculation, do not directly characterize extreme event risk. Some investments to improve reliability using reliability metrics may also improve resilience, but this is not quantifiable unless resilience metrics are also used.

**QUANTIFYING INVESTMENT BENEFITS BY RERUNNING HISTORY**

In this section, we explain the historical rerun method used to measure the resilience benefits of upgrading infrastructure to withstand higher wind speeds. We first assess the historical resilience performance of the system using the risk metrics explained earlier. We then consider what would have happened if a specific investment to upgrade the system had been made several years ago. How would that investment have improved the system's performance over the years? Based on this concept, we modify the outage data to reflect the effects of such an investment made several years back and then recalculate the risk metrics using the modified outage data. Comparing the results with and without the investment quantifies the impact of that investment. The historical rerun method is discussed in more detail in [5].

The historical rerun method is limited in that while it can maintain or decrease the number of outages, it cannot synthesize new outages. Nevertheless, it could still be used to model the effects of an increased severity of events, such as an average increase in wind speeds, as long as there is sufficient proposed hardening to result in an overall decrease in the outage rate. Moreover, an expected future increase in the frequency of extreme events is easily accommodated by increasing the average annual event rate $\bar{r}_{event}$.

**Modeling the benefits of wind hardening**

We develop the area outage rate curve [5] using the outage data and weather data. The area outage rate curve gives the empirical average outage rate of an area of the distribution system as a function of wind speed. The area outage rate curve for our utility data can be fit by an exponential function [5]. Wind hardening has the effect of shifting the area outage rate curve to the left, i.e., we see lower outage rates at each wind speed level as compared to the outage rates before the upgrade. We shift the area outage rate curve to the left to represent a wind-hardening investment for a 10% decrease in outage rates. 2000 samples are taken randomly from the outage data according to the reduced outage rates, and the risk metrics are computed for each sample, and then averaged over the 2000 samples.

**Modeling the benefits of faster restoration**

While investments can be made for hardening the infrastructure, investments can also be made to improve the restoration of outages. We model such investments as well using the historical rerun technique. If investments had been made to acquire more repair crews, better stocks of spare parts, and better route scheduling, then the restoration rates of the outage events would have improved, resulting in the earlier completion of the restorations. To demonstrate the effects of such an investment on the risk metrics, we assume investments that would have resulted in a 10% faster restoration and rerun history by updating the restoration times of outages in the data accordingly.

## RESULTS OF RESILIENCE INVESTMENTS

The effects of wind hardening are shown in Table 1. As a result of the 10% wind hardening investment, there is an almost 12% less chance that an outage event would have cost more than 3.3×10⁶ USD. Since the wind hardening decreases the number of outages in general, the annual event rate also sees a 6% decrease. Consequently, the expected annual frequency of large-cost events also decreases by nearly 17%.

*Table 1: The effects of historical rerun with 10% wind hardening*

| Metric | Before | After | % diff. |
|---|---|---|---|
| $\alpha$ | 0.789 | 0.792 | 0.1% |
| $p_{large}$ | 0.010 | 0.009 | -11.6% |
| $\bar{r}_{event}$ | 618 | 581 | -6.0% |
| $f_{large}$ | 6.2 | 5.1 | -16.9% |

The effects of faster restoration are shown in Table 2. Investments made for improving the restoration rate of outage events by 10% would have resulted in a significant 22% decrease in the probability of large cost events. In other words, there would have been 22% less chance that an outage event costing 3.3×10⁶ USD or higher would occur if such an investment had been made. This is also reflected in the average annual frequency of the large cost events also decreasing by almost 22%. The number of events and their average annual rate $\bar{r}_{event}$ are unchanged by the faster restoration.

*Table 2: The effects of historical rerun with 10% faster restoration*

| Metric | Before | After | % diff. |
|---|---|---|---|
| $\alpha$ | 0.789 | 0.821 | 1.8% |
| $p_{large}$ | 0.010 | 0.008 | -21.6% |
| $f_{large}$ | 6.2 | 4.8 | -21.6% |

An important detail about the faster restoration modeling is that outage events with only one outage are not affected by it as their restoration process, which starts with the first restore, has zero duration. There are 2142 such events in the utility data, and thus, their costs remain the same after the faster restoration modeling.

## CONCLUSIONS

Fundamental to our analysis of distribution system utility data is grouping observed outages into events in which outages accumulate before they are restored. Events of all sizes are easily extracted from utility outage data [2]. It is straightforward to evaluate the customer hours lost for each event. Then we calculate a cost for each event that is proportional to the customer hours.

We define risk by an event cost exceedance curve [7]; that is, the probability that the cost of an event exceeds a given amount. We define large cost events as events with costs exceeding a threshold value $c_{large}$. In particular, the probability of large cost events $p_{large}$ is the probability that an event has cost greater than $c_{large}$ as well as the value of the cost exceedance curve at $c_{large}$. For the larger costs in our utility data, the cost exceedance curve is linear with slope magnitude $\alpha$ on a log-log plot. It follows that the event cost exceedance curve above $c_{large}$ and hence the risk of large events is described by its value $p_{large}$ together with the slope magnitude $\alpha$. We propose the probability of large cost events

$p_{large}$ and the slope magnitude $\alpha$ as novel large event risk metrics. This new formulation of extreme event risk in distribution systems incorporates both the probability and cost of extreme events.

Our utility data shows a cost exceedance curve with a heavy tail, showing that large cost events will occasionally happen, and have substantial risk. Further, the heaviness of the tail (slope magnitude $\alpha < 1$) and the consequent expected occurrence of catastrophic events with extremely large costs raises questions about using average or mean values to characterize large costs in distribution systems that should be addressed in future work.

We use the historical rerun method to quantify the resilience improvement that a proposed resilience investment would have had if the investment had been made in the past. This method has several advantages:

- It is driven by real historical data, which includes all the factors that are very difficult to capture in models, such as weather, human factors, emergency system reconfigurations, equipment aging, and restoration practices.
- Model based approaches, often used for predicting future behaviors, have considerable uncertainties such as modeling errors. However, using historical data removes many of these uncertainties.
- More frequent wind events and to a limited extent more severe wind events could be accommodated.
- It is easy to communicate the benefits of the proposed resilience investment to stakeholders, as those benefits would have applied to the already lived experience of stakeholders in the past, particularly for large events.
- It is computationally inexpensive and easy to implement, and the outage data is already available to utilities.

Future work will include analyzing outage and cost data from additional distribution systems, extending the analysis to utility costs, and quantifying other resilience investments such as undergrounding.

**Acknowledgments**

Support from Iowa State University EPRC, PSerc, NSF grant 2153163, and DOE is gratefully acknowledged.